\documentclass[12pt]{article}
\usepackage{amsmath}

\usepackage{doublespace}

\input{tcilatex}

\begin{document}

\bigskip 

\bigskip

\begin{center}
\bigskip \textbf{A note on estimating stochastic volatility and its
volatility: a new simple method}

\bigskip
\end{center}

\bigskip

\bigskip

With the exception of Alghalith (2012), the previous methods of estimating
(stochastic) volatility suffered two major limitations. Firstly, they are
only applicable to \textit{time-series} data and therefore the volatility of
the investor's portfolio (a cross section of assets) for time $t$ cannot be
properly and directly estimated. Secondly, these methods required the
arbitrary generation of data series for volatility in order to estimate the
volatility. This seems somewhat contradictory and futile. Alghalith (2012) \
used a stochastic factor model, where volatility is a function of an
external economic factor (GDP). In this note, we use a \textit{stochastic}
volatility (in the sense of Heston, etc.). In doing so, we overcome the
aforementioned limitations. Below is a description of the theoretical model.

We modify the standard portfolio model to include a stochastic volatility.
We still use a two-dimensional Brownian motion $\left\{ W_{1s},W_{2s},%
\mathcal{F}_{s}\right\} _{t\leq s\leq T}$ on the probability space $\left(
\Omega ,\mathcal{F}_{s},P\right) ,$ where $\left\{ \mathcal{F}_{s}\right\}
_{t\leq s\leq T}$ is the augmentation of filtration. We include a risky
asset and a risk-free asset. The risk-free asset price process is given by $%
S_{0}=e^{\int\limits_{t}^{T}rds},$ where $r\in C_{b}^{2}\left( R\right) $ is
the rate of return.

The dynamics of the risky asset price are given by

\begin{equation}
dS_{s}=S_{s}\left\{ \mu _{s}ds+\sigma _{s}dW_{1s}\right\} ,
\end{equation}%
where $\mu _{s}$ and $\sigma _{s}$ are the rate of return and the
volatility, respectively.

As in Heston's model, the stochastic volatility is given by 
\begin{equation}
d\sigma _{s}^{2}=\left( \alpha -\beta \sigma _{s}^{2}\right) ds+\gamma
\sigma _{s}dW_{2s},\sigma _{t}^{2}\equiv \bar{\sigma},
\end{equation}%
where $\left| \rho \right| <1$ is the correlation factor between the two
Brownian motions, $\alpha ,$ $\beta $ and $\gamma $ are constants.

The wealth process is given by

\begin{equation}
X_{T}^{\pi }=x+\int\limits_{t}^{T}\left\{ r_{s}X_{s}^{\pi }+\left( \mu
_{s}-r_{s}\right) \pi _{s}\right\} ds+\int\limits_{t}^{T}\pi _{s}\sigma
_{s}dW_{1s},
\end{equation}%
where $x$ is the initial wealth, $\left\{ \pi _{s},\mathcal{F}_{s}\right\}
_{t\leq s\leq T}$ is the portfolio process, and $E\int\limits_{t}^{T}\pi
_{s}^{2}ds<\infty .$ The trading strategy $\pi _{s}\in \mathcal{A}\left( x,%
\bar{\sigma}\right) $ is admissible$.$ $X_{s}$ $=\pi _{s}+B_{s},$where $%
B_{s} $ is the amount invested in the risk-free asset.

\bigskip The investor's objective is to maximize the expected utility of the
terminal wealth

\[
V\left( t,x,\bar{\sigma}\right) \text{ =}\underset{\pi }{\sup }E\left[
U\left( X_{T}^{\pi }\right) \mid \mathcal{F}_{t}\right] ,
\]%
where $V\left( .\right) $ is the value function, $U\left( .\right) $ is
continuous, smooth, bounded and strictly concave utility function.

The value function satisfies the Hamilton-Jacobi-Bellman PDE (suppressing
the notations)%
\[
V_{t}+rxV_{x}+\left( \alpha -\beta \bar{\sigma}\right) V_{\bar{\sigma}}+%
\frac{1}{2}\gamma ^{2}\bar{\sigma}V_{\bar{\sigma}\bar{\sigma}}+
\]%
\[
\underset{\pi }{Sup}\left\{ \frac{1}{2}\pi ^{2}\bar{\sigma}V_{xx}+\left[ \pi
\left( \mu -r\right) \right] V_{x}+\gamma \rho \bar{\sigma}\pi V_{x\bar{%
\sigma}}\right\} =0,
\]%
\[
V\left( T,x,\bar{\sigma}\right) =u\left( x,\bar{\sigma}\right) .
\]%
Thus the solution yields

\begin{equation}
\pi ^{\ast }=-\frac{\left( \mu -r\right) V_{x}}{\bar{\sigma}V_{xx}}-\rho
\gamma \frac{V_{x\bar{\sigma}}}{V_{xx}}.  \label{20}
\end{equation}

We take a second-order Taylor's expansion of $V\left( t,x,\bar{\sigma}%
\right) $ and therefore 
\[
V_{x}\left( t,x,\bar{\sigma}\right) \approx \alpha _{0}+\alpha _{1}x+\alpha
_{2}\bar{\sigma}.
\]%
Substituting this into $\left( \ref{20}\right) $ yields

\[
\pi ^{\ast }=-\frac{\left( \mu -r\right) \left[ \alpha _{0}+\alpha _{1}(\pi
^{\ast }+B)+\alpha _{2}\bar{\sigma}\right] }{\alpha _{1}\bar{\sigma}}-\rho
\gamma \frac{\alpha _{2}}{\alpha _{1}}.
\]%
The above equation can be rewritten as%
\begin{equation}
\pi ^{\ast }=\beta _{2}-\frac{\left( \mu -r\right) \pi ^{\ast }}{\bar{\sigma}%
}+\frac{\beta _{1}\left( \mu -r\right) }{\bar{\sigma}};\beta _{1}\equiv -%
\frac{\alpha _{0}+\alpha _{1}B+\alpha _{2}\bar{\sigma}}{\alpha _{1}},\beta
_{2}\equiv -\rho \gamma \frac{\alpha _{2}}{\alpha _{1}},
\end{equation}%
where $\beta _{i}$ is a constant. Thus

\[
\pi ^{\ast }=\frac{\beta _{2}}{1+\frac{\mu -r}{\bar{\sigma}}}+\frac{\beta
_{1}\left( \mu -r\right) }{\bar{\sigma}+\mu -r}.
\]%
The above equation can be rewritten as the following regression equations%
\begin{equation}
\pi _{t}^{\ast }=\frac{\beta _{2}}{1+\frac{\mu -r}{\beta _{3}}}+\frac{\beta
_{1}\left( \mu -r\right) }{\beta _{3}+\mu -r},  \label{21}
\end{equation}%
where $\beta _{i}$ is a parameter to be estimated by a non-linear regression
(while the variables $\pi ^{\ast },$ $\mu $ and $r$ are observed data), and $%
\beta _{3}$ is an estimate of the volatility of the portfolio for period $t.$

To estimate the volatility of volatility $\gamma ,$ we multiply $\left( \ref%
{21}\right) $ by $\gamma $ and take the inverse to obtain 
\[
\breve{\pi}=\frac{\beta _{4}}{\frac{\beta _{5}}{1+\frac{\mu -r}{\hat{\beta}%
_{3}}}+\frac{\beta _{6}\left( \mu -r\right) }{\hat{\beta}_{3}+\mu -r}},\beta
_{4}\equiv \gamma ,\breve{\pi}\equiv \frac{1}{\pi _{t}^{\ast }},
\]%
where $\beta _{4}$ is an estimate of $\gamma ,$ $\hat{\beta}_{3}$ is the
estimated volatility from the previous regression, and $\breve{\pi}$ is
observed data. A similar procedure can be used to estimate the factor of
correlation.

\end{document}